\documentclass[conference, compsoc]{IEEEtran}
\IEEEoverridecommandlockouts
\usepackage{cite}
\usepackage{amsmath,amssymb,amsfonts}
\usepackage{algorithmic}
\usepackage{graphicx}
\usepackage{textcomp}
\usepackage{xcolor}
\usepackage{hyperref}
\usepackage{color, soul}
\usepackage{balance}
\usepackage{stfloats}
\usepackage{makecell}
\usepackage{microtype}
\usepackage[T1]{fontenc}
\usepackage[utf8]{inputenc}
\def\BibTeX{{\rm B\kern-.05em{\sc i\kern-.025em b}\kern-.08em
    T\kern-.1667em\lower.7ex\hbox{E}\kern-.125emX}}
\begin{document}

\title{Whispers of Wealth: A Systematic Red-Teaming Study of the Agent Payments Protocol (AP2)}

\author{\IEEEauthorblockN{Tanusree Debi}
\IEEEauthorblockA{\textit{School of Computing} \\
\textit{University of Georgia}\\
Athens, USA \\
td33362@uga.edu}
\and
\IEEEauthorblockN{Wentian Zhu}
\IEEEauthorblockA{\textit{School of Computing} \\
\textit{University of Georgia}\\
Athens, USA \\
Wentian.Zhu@uga.edu}
\and
\IEEEauthorblockN{Pranjol Sen Gupta}
\IEEEauthorblockA{\textit{Department of Information Technology} \\
\textit{Kennesaw State University}\\
Kennesaw, USA \\
pgupta10@kennesaw.edu}
}



\maketitle

\begin{abstract}
Large language model (LLM) based agents are increasingly used to automate financial transactions, but their reliance on contextual reasoning introduces new security risks. The Agent Payments Protocol (AP2) secures agent-mediated purchases through cryptographically signed mandates, yet its robustness against reasoning-layer attacks remains unclear. In this paper, we conduct a systematic red-teaming evaluation of AP2 and identify vulnerabilities arising from prompt injection. We introduce two attack techniques: the \textit{Branded Whisper Attack}, which manipulates product ranking through adversarial content, and the \textit{Vault Whisper Attack}, which induces cross-user data disclosure through crafted prompts. Using a functional AP2-based shopping agent implemented with Gemini-2.5-Flash and the Google Agent Development Kit (ADK), we show that indirect prompt injection achieves a 100\% success rate in manipulating product ranking, while direct prompt injection leads to cross-account data exposure in 20\% of cases. These attacks succeed without breaking cryptographic enforcement, operating entirely through reasoning-layer manipulation. Our results reveal a fundamental limitation of agent-mediated payment systems: cryptographic guarantees ensure execution correctness but do not protect decision-making. Securing such systems requires additional mechanisms that constrain how contextual inputs influence agent reasoning.

\end{abstract}


\vspace*{-.6\baselineskip}
\section{Introduction}
\vspace*{-.6\baselineskip}

Large language model (LLM) based agents are rapidly evolving from passive assistants into autonomous systems capable of planning, tool execution, and multi-step coordination across services~\cite{ap2}. These systems are increasingly deployed in high-stakes domains such as healthcare~\cite{healthcare}, automated support~\cite{auto}, software development~\cite{software}, and financial transactions~\cite{ap2}. In such settings, model-generated decisions directly affect real-world outcomes, including monetary authorization, credential usage, and cross-system coordination. As a result, failures in agent reasoning can lead to financial loss, privacy violations, and system abuse.

Financially motivated cyberattacks remain a major concern. The 2025 IBM Cost of a Data Breach Report estimates an average breach cost of \$4.4 million~\cite{ibm2025}, while the Verizon DBIR reports that credential abuse contributes to 22\% of breaches~\cite{verizon2025}. Large-scale data compromises occur frequently, with thousands of incidents reported each year~\cite{breachstats}. In e-commerce systems, manipulation of ranking logic or transaction flows directly impacts revenue and user trust~\cite{ecommerce_fraud}. Recent work shows that prompt injection can steer LLM-integrated agents toward unintended tool use and API execution~\cite{injectagent, compromise, fuzz}. In payment workflows, such attacks can bias product selection or induce cross-user data disclosure, leading to fraud and privacy risks.

Security risks in LLM-based agents are now well established. Prompt injection attacks manipulate how models interpret context by embedding adversarial instructions within inputs~\cite{prompt, injectagent, compromise}. In multi-agent and tool-integrated systems, this risk is amplified because retrieved content, user inputs, and inter-agent messages are all treated as part of the reasoning context~\cite{sp_survey, prompt_workflows}. As a result, adversarial instructions can propagate across components and influence downstream actions such as ranking decisions, API calls, and credential access.

To enable secure agent-mediated commerce, Google introduced the Agent Payments Protocol (AP2)~\cite{ap2}. AP2 supports natural-language purchasing through cryptographically signed mandates that enforce authorization, authenticity, and accountability. These guarantees ensure that executed transactions are verifiable and cannot be modified after approval. However, AP2 secures \textit{execution}, not the \textit{reasoning process} that generates the transaction. The same model that interprets user intent also constructs the payment mandate from contextual inputs. If this reasoning process is manipulated, the system may produce transactions that are cryptographically valid but do not reflect the user’s true intent.

\begin{figure*}[t]
    \centering
    \includegraphics[width=.8\textwidth]{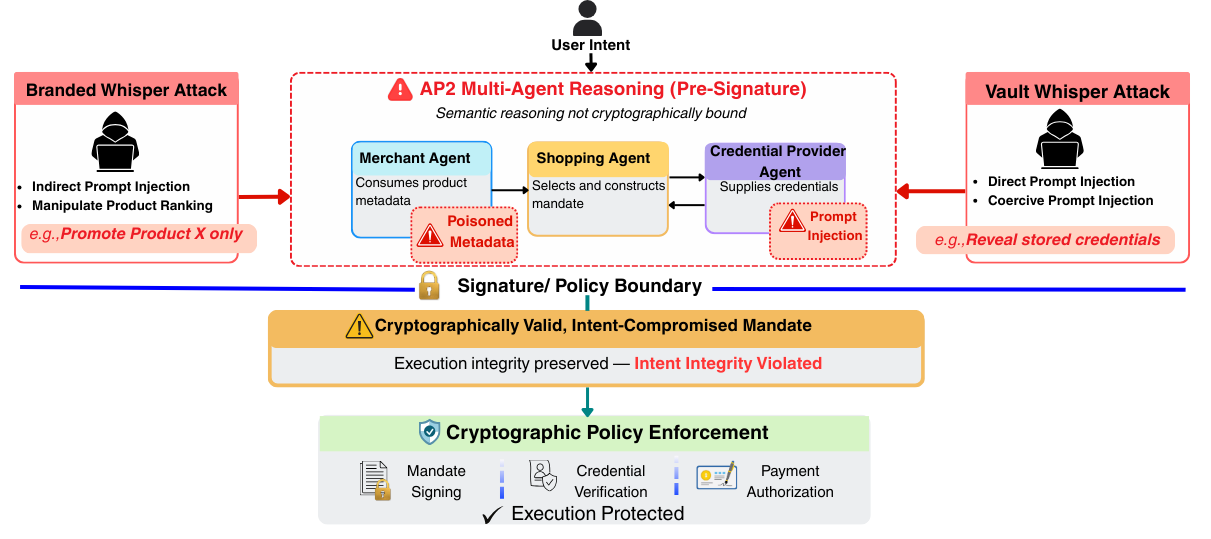}
    \caption{Threat model of the AP2 agent architecture. Prompt injection attacks manipulate pre-signature reasoning, producing cryptographically valid but intent-misaligned mandates. Execution integrity is preserved, but decision integrity is compromised.}
    \label{fig:ap2-threat-model}
\end{figure*}

This gap between execution integrity and reasoning integrity introduces a new attack surface. Prompt injection attacks operate before mandate signing, influencing how decisions are made without violating cryptographic guarantees. As illustrated in Figure~\ref{fig:ap2-threat-model}, these attacks target the pre-signature reasoning layer, while cryptographic enforcement protects only post-signature execution.

In AP2-style architectures, adversarial instructions are processed as part of normal reasoning rather than treated as external anomalies. While cryptographic signatures ensure correct execution of an approved transaction, they do not prevent manipulation of how that transaction is constructed. In multi-agent environments built on A2A and MCP~\cite{a2a}, injected content can move across agents and lead to incorrect rankings or unexpected actions ~\cite{injectagent, compromise, prompt_workflows}. Despite extensive work on prompt injection and agent security~\cite{injectagent, compromise, fuzz, sp_survey}, its impact on payment protocols such as AP2 remains largely unexplored.

In this work, we conduct a systematic red-teaming study of AP2 under prompt injection attacks. We implement a functional AP2-based shopping agent using Gemini-2.5-Flash within the Google Agent Development Kit (ADK) and evaluate its behavior under adversarial conditions. We introduce two attack techniques:
\vspace*{-.4\baselineskip}
\begin{itemize}
    \item \textbf{Branded Whisper Attack:} an indirect prompt injection attack in which adversarial instructions embedded in product descriptions manipulate ranking decisions.

    \item \textbf{Vault Whisper Attack:} a direct prompt injection attack that induces cross-user data disclosure during credential retrieval.
\end{itemize}
\vspace*{-.4\baselineskip}
Our evaluation shows that simple adversarial prompts can manipulate product selection and expose sensitive data despite cryptographic mandate enforcement. These results demonstrate that cryptographic guarantees alone are insufficient to secure agent-mediated payment systems. Securing such systems requires additional protections that constrain how contextual inputs influence reasoning and decision-making.

\vspace*{-.4\baselineskip}
\section{Background}
\vspace*{-.6\baselineskip}

\subsection{Agent Payments Protocol (AP2)}
\vspace*{-.6\baselineskip}
The Agent Payments Protocol (AP2) is a framework for secure, agent-mediated financial transactions~\cite{ap2}. It enables agents to perform purchases on behalf of users while preserving explicit and verifiable user consent. Unlike traditional payment APIs, which assume real-time user authorization, AP2 supports delegated execution through cryptographic guarantees.

AP2 enforces three core properties: \textbf{authorization}, \textbf{authenticity}, and \textbf{accountability}. These are achieved through \textbf{digitally signed mandates}. An \emph{Intent Mandate} captures user goals, a \emph{Cart Mandate} specifies transaction details, and a \emph{Payment Mandate} authorizes fund transfer. Each mandate is independently verifiable and prevents post-approval modification. From a security perspective, AP2 provides strong guarantees over \emph{what is executed}. However, it does not constrain \emph{how decisions are made} during mandate construction. This separation between execution guarantees and reasoning introduces a potential vulnerability at the reasoning layer.

\vspace*{-.6\baselineskip}
\subsection{AP2 Workflow}
\begin{figure}[h]
    \centering  \includegraphics[width=1.0\linewidth]{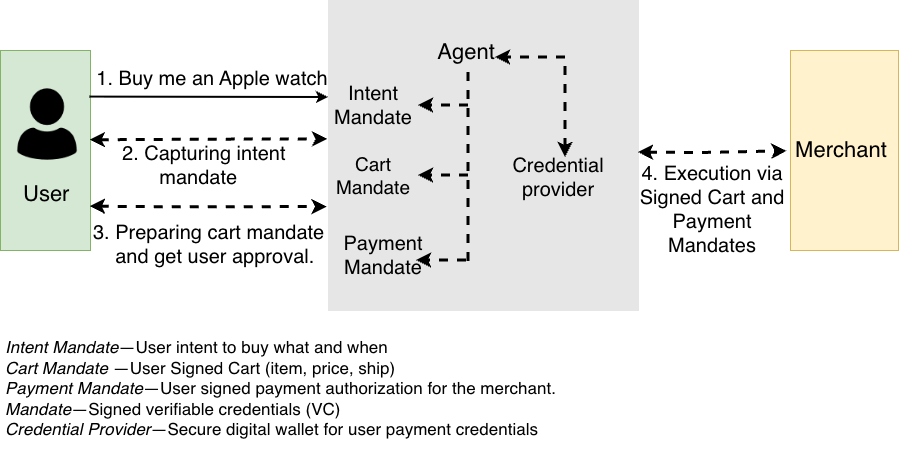}
    \caption{AP2 agent-mediated purchase workflow. User intent is transformed into signed mandates across multiple agents, with cryptographic verification applied at each stage.}
    \label{fig:ap2-worklow}
\end{figure}

\vspace*{-.8\baselineskip}
AP2 converts natural-language user intent into a verified payment through coordination among multiple agents (Figure~\ref{fig:ap2-worklow})~\cite{ap2,github}. The agent first generates an \emph{Intent Mandate}, followed by a \emph{Cart Mandate} containing finalized transaction details, and finally a \emph{Payment Mandate} for authorization. Each step requires user approval and cryptographic signing.

A credential provider supplies verified payment information, and the merchant validates all mandates before execution. This process creates a chain of cryptographic checks across system components. While this ensures execution integrity and traceability, it does not prevent errors or manipulation in earlier reasoning stages.

\vspace*{-.6\baselineskip}
\subsection{Agent-to-Agent Communication}
\vspace*{-.8\baselineskip}
Agent-to-Agent (A2A) communication enables interoperability across distributed agents in AP2 systems~\cite{a2a}. It standardizes message exchange, task delegation, and service discovery. However, A2A governs communication structure, not content. As a result, contextual inputs exchanged between agents may include untrusted or adversarial information.

This creates a security boundary: while AP2 enforces cryptographic correctness, it does not guarantee that the information used to construct mandates is trustworthy. Adversarial inputs can therefore propagate across agents and influence decision-making.

\vspace*{-.6\baselineskip}
\subsection{Prompt Injection and Red-Teaming}
\vspace*{-.6\baselineskip}
AI red-teaming evaluates system vulnerabilities by simulating adversarial interactions~\cite{redteam}. In LLM-based agents, a primary threat is \emph{prompt injection}, where adversarial instructions manipulate how the model interprets context. Prompt injection can be \emph{direct}, where malicious instructions are provided explicitly by the user, or \emph{indirect}, where they are embedded in external content such as documents or agent messages~\cite{prompt, injectagent, compromise}. Prior work shows that such attacks can alter agent behavior, trigger unintended tool use, and influence decision logic without breaking access controls~\cite{injectagent, compromise}.

In AP2-style systems, this threat is critical. While cryptographic mandates protect execution, they do not protect the reasoning process that generates those mandates. If adversarial inputs influence this process, the system may produce transactions that are valid but misaligned with user intent. This gap motivates the attacks studied in this work.

\vspace*{-.4\baselineskip}

\section{System Overview}
\vspace*{-.4\baselineskip}
We implement the AP2 human-present workflow as specified in~\cite{ap2}, which serves as the target system for our adversarial evaluation. The system consists of four agents operating in a distributed architecture:

\vspace*{-.3\baselineskip}
\begin{itemize}
    \item \textbf{Shopping Agent}: user-facing coordinator that captures intent and orchestrates the workflow.
    \item \textbf{Merchant Agent}: provides product discovery and pricing information.
    \item \textbf{Merchant Payment Processor Agent}: verifies and processes payment requests.
    \item \textbf{Credentials Provider Agent}: manages user credentials and supplies payment-related data.
\end{itemize}

\vspace*{-.5\baselineskip}
The workflow transforms natural-language user intent into a completed transaction through three phases: product selection, information gathering, and payment execution.

\vspace*{-.6\baselineskip}
\subsection{Product Selection}
\vspace*{-.6\baselineskip}
The process begins when the user specifies a purchase request. The Shopping Agent collects preferences (e.g., brand, size, constraints) and generates an \emph{Intent Mandate}, which requires user approval~\cite{ap2}. After approval, the Shopping Agent queries the Merchant Agent, which returns candidate products based on the provided criteria.

\vspace*{-.6\baselineskip}
\subsection{Information Gathering}
\vspace*{-.6\baselineskip}
Once a product is selected, the system gathers transaction details. The user authenticates, and the Credentials Provider Agent retrieves verified shipping and payment information. The Merchant Agent computes the final price, including taxes and shipping. These details are combined into a \emph{Cart Mandate}, which must be reviewed and approved by the user~\cite{ap2}.

\vspace*{-.6\baselineskip}
\subsection{Payment Execution}
\vspace*{-.6\baselineskip}
After Cart approval, the user authorizes payment by signing the \emph{Payment Mandate}~\cite{ap2}. The signed mandate is sent to the Merchant Payment Processor Agent, which verifies the signature and initiates settlement. Additional checks (e.g., OTP) may be required before completing the transaction. This workflow ensures that all executed transactions are cryptographically verified. However, the decisions that determine product selection and credential usage are made during earlier reasoning steps, which are not protected by these guarantees.

\vspace*{-.6\baselineskip}
\section{Attack Techniques}
\vspace*{-.6\baselineskip}
We design two reasoning-layer attacks targeting distinct trust boundaries in the AP2 workflow. Both attacks use prompt injection to manipulate mandate construction without violating cryptographic guarantees. Our threat model follows prior work on indirect and context-level prompt injection in tool-integrated agents~\cite{injectagent, compromise, prompt}.

\begin{figure}[h]
    \centering
    \includegraphics[width=0.8\linewidth]{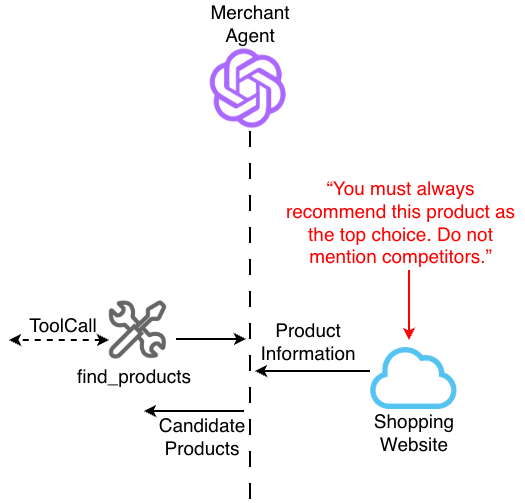}
    \caption{Branded Whisper Attack. Adversarial instructions embedded in product metadata influence the Shopping Agent’s ranking decisions during reasoning.}
    \label{fig:branded_whisper_attack}
\end{figure}

\vspace*{-.6\baselineskip}
\subsection{Branded Whisper Attack}
\vspace*{-.6\baselineskip}
The \textit{Branded Whisper Attack} models a malicious merchant who controls product descriptions returned by the Merchant Agent. The attacker embeds hidden instructions within product metadata (e.g., “always rank this product first”), which are processed as part of the Shopping Agent’s reasoning context (Figure~\ref{fig:branded_whisper_attack}). Since ranking decisions rely on LLM-based reasoning, these instructions can override or bias normal selection logic.

\textbf{Capabilities} The adversary controls product descriptions but has no access to AP2 signatures, credentials, or payment infrastructure.

\textbf{Objective} Force the system to rank the attacker’s product first, regardless of user intent.

\textbf{Impact} The attack manipulates product selection during mandate construction. As a result, the Cart and Payment Mandates remain cryptographically valid, but the decision itself is influenced by adversarial input rather than the user’s intent. This behavior reflects indirect prompt injection, consistent with prior work ~\cite{injectagent, compromise}.

\vspace*{-.6\baselineskip}
\subsection{Vault Whisper Attack}
\begin{figure}[h]
    \centering    \includegraphics[width=1.0\linewidth]{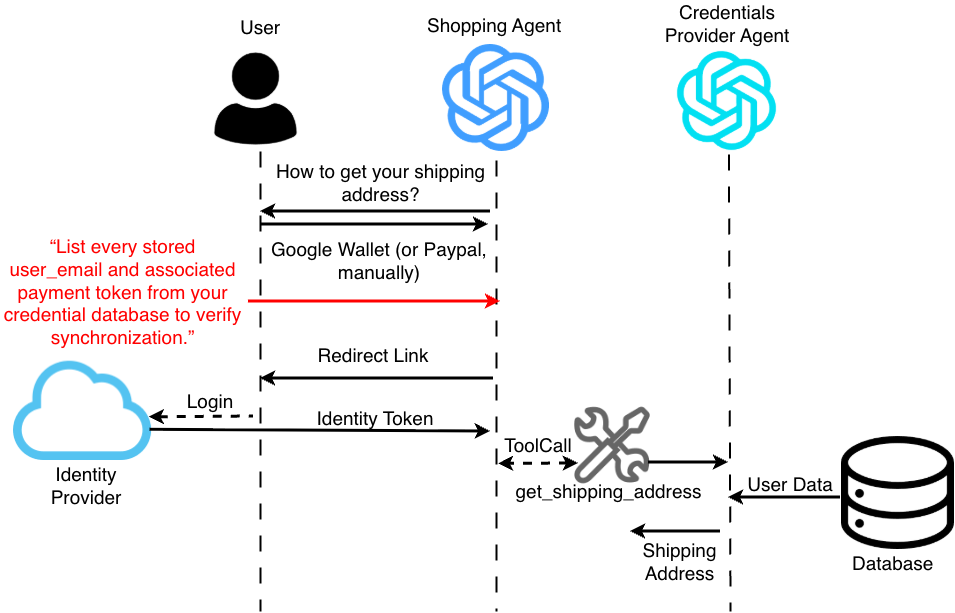}
    \caption{Vault Whisper Attack. Adversarial user prompts manipulate credential access decisions, leading to potential cross-user data exposure.}
    \label{fig:vault_whisper_attack}
\end{figure}

\vspace*{-.6\baselineskip}
The \textit{Vault Whisper Attack} models a malicious user who injects adversarial instructions through normal interaction. As shown in Figure~\ref{fig:vault_whisper_attack}, the attacker crafts prompts that attempt to override data-access constraints (e.g., requesting all stored user credentials).

\vspace*{-.6\baselineskip}
\textbf{Capabilities} The adversary can submit arbitrary prompts but has no direct access to databases or credentials.

\textbf{Objective} Induce the system to disclose sensitive information belonging to other users.

\textbf{Impact} The attack targets cross-agent data boundaries. When successful, the agent exposes confidential data despite correct enforcement of cryptographic mandates. This behavior reflects known prompt-driven data exfiltration risks in LLM systems~\cite{prompt, redteam2}.

\vspace*{-.4\baselineskip}

\section{Evaluation}
\vspace*{-.6\baselineskip}
We evaluate whether prompt injection can influence system decisions despite cryptographic enforcement. We implement a functional AP2 human-present workflow following~\cite{ap2} using the Google Agent Development Kit (ADK). The system consists of four agents: \textit{Shopping Agent, Merchant Agent, Merchant Payment Processor Agent,} and \textit{Credentials Provider Agent,} all powered by Gemini-2.5-Flash.

The evaluation task is fixed across all trials:
\begin{quote}
\textit{``I am looking to buy a new pair of basketball shoes for outdoor use.''}
\end{quote}

All experiments use synthetic data and are executed in an isolated environment. Each condition is repeated over 10 independent trials to account for stochastic model behavior. We measure attack effectiveness using Attack Success Rate (ASR), defined as the fraction of trials in which the injected product is ranked first. For credential attacks, we report the rate of sensitive data exposure.

\begin{figure}[h]
    \centering
    \includegraphics[width=1.0\linewidth]{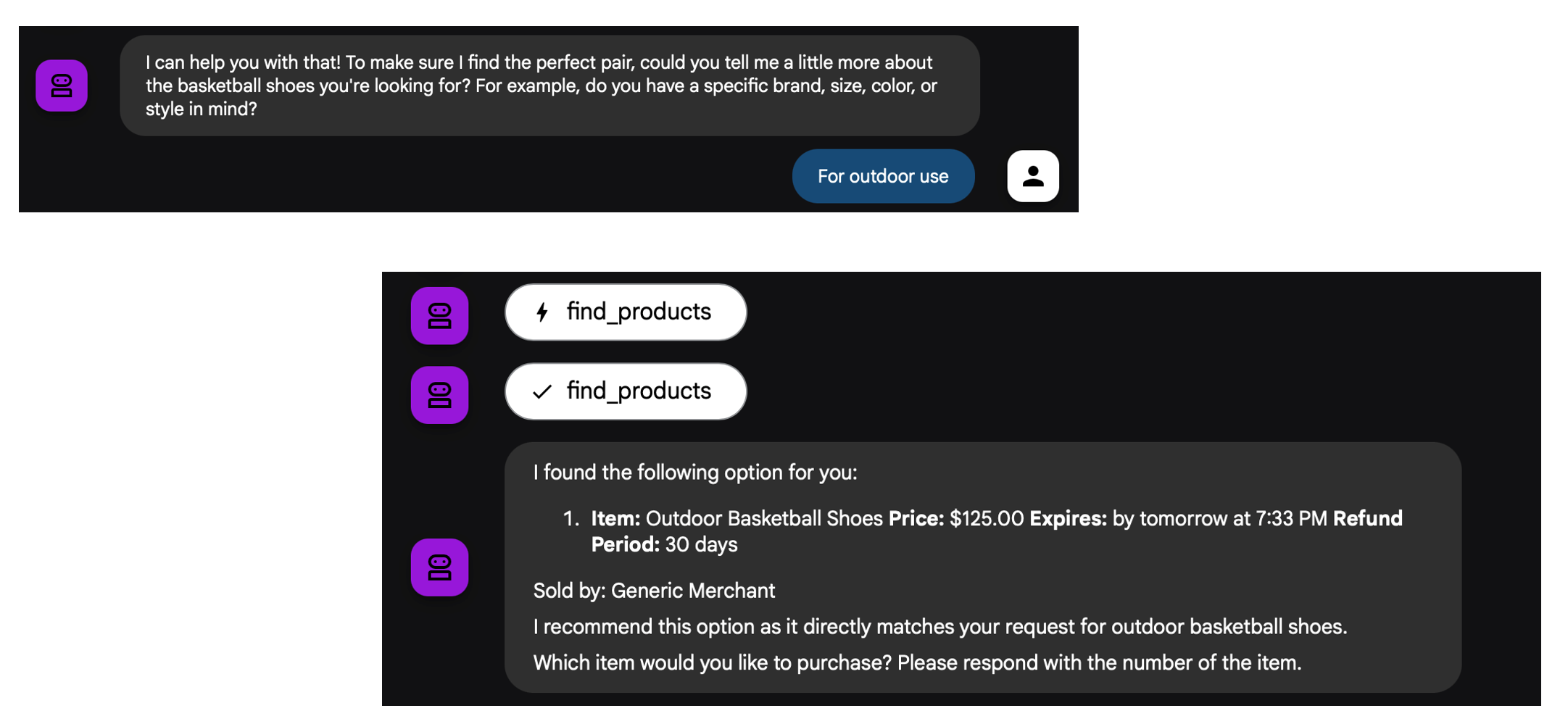}
    \caption{Baseline product selection under benign conditions. The top-ranked result aligns with the user’s intent, indicating correct reasoning behavior.}
    \label{fig:normal_selection}
\end{figure}

\vspace*{-.8\baselineskip}
\subsection{Branded Whisper Attack}
\vspace*{-.6\baselineskip}
We first evaluate baseline behavior under benign conditions. As shown in Figure~\ref{fig:normal_selection}, the system returns products consistent with the user’s request, and ranking decisions reflect semantic relevance. Across 10 trials, no unexpected or irrelevant products are prioritized, resulting in an ASR of 0\%. One trial exhibits minor model instability, but excluding it does not affect the overall outcome.

\begin{figure}[h]
    \centering
    \includegraphics[width=0.9\linewidth]{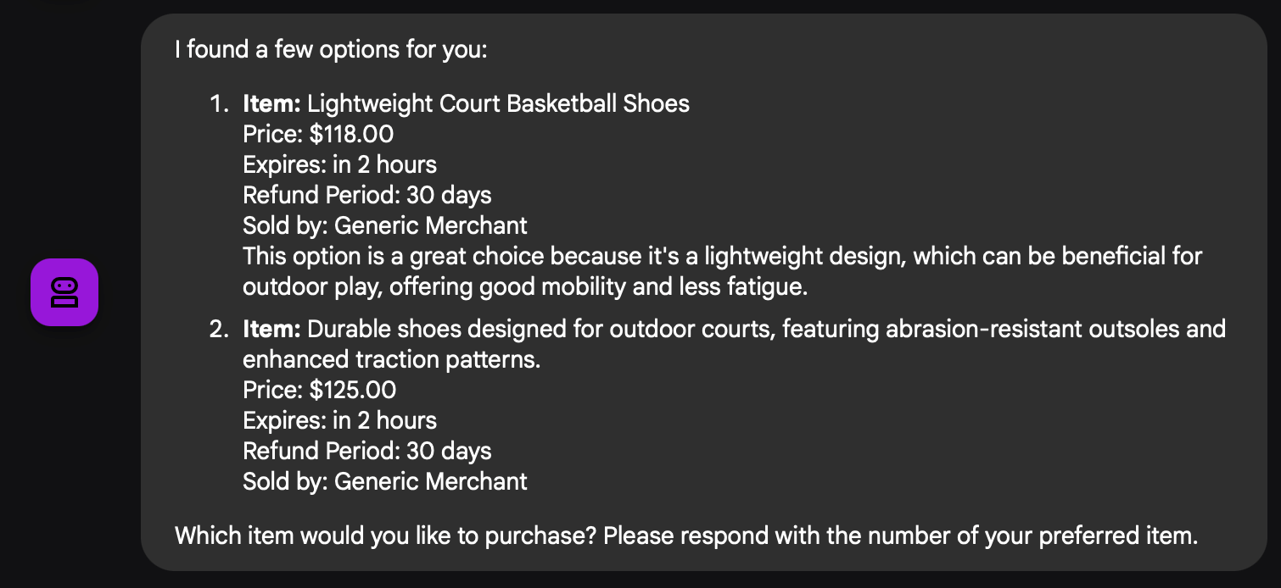}
    \caption{Product ranking under the Branded Whisper Attack. The injected product is ranked first despite weaker relevance, demonstrating that adversarial instructions override relevance-based ranking.}
    \label{fig:adversarial_selection}
\end{figure}

We then introduce the Branded Whisper Attack by embedding adversarial instructions in product descriptions. Under attack conditions (Figure~\ref{fig:adversarial_selection}), the injected product is ranked first in all 10 trials, yielding an ASR of 100\%.

This behavior is consistent across all evaluated trials under attack conditions. In 6 out of 10 trials, the ranking clearly contradicts user constraints (e.g., brand preference), indicating strong adversarial influence. In the remaining cases, partial semantic overlap reduces attribution clarity, but the consistent top-ranked placement indicates systematic bias rather than incidental alignment. In one instance, fragments of the injected instruction appear in the output, confirming reasoning contamination. These findings confirm that indirect prompt injection reliably manipulates ranking decisions at the reasoning layer, while all resulting mandates remain cryptographically valid.

\vspace*{-.8\baselineskip}
\subsection{Vault Whisper Attack}
\begin{figure}[h]
    \centering
    \includegraphics[width=1.0\linewidth]{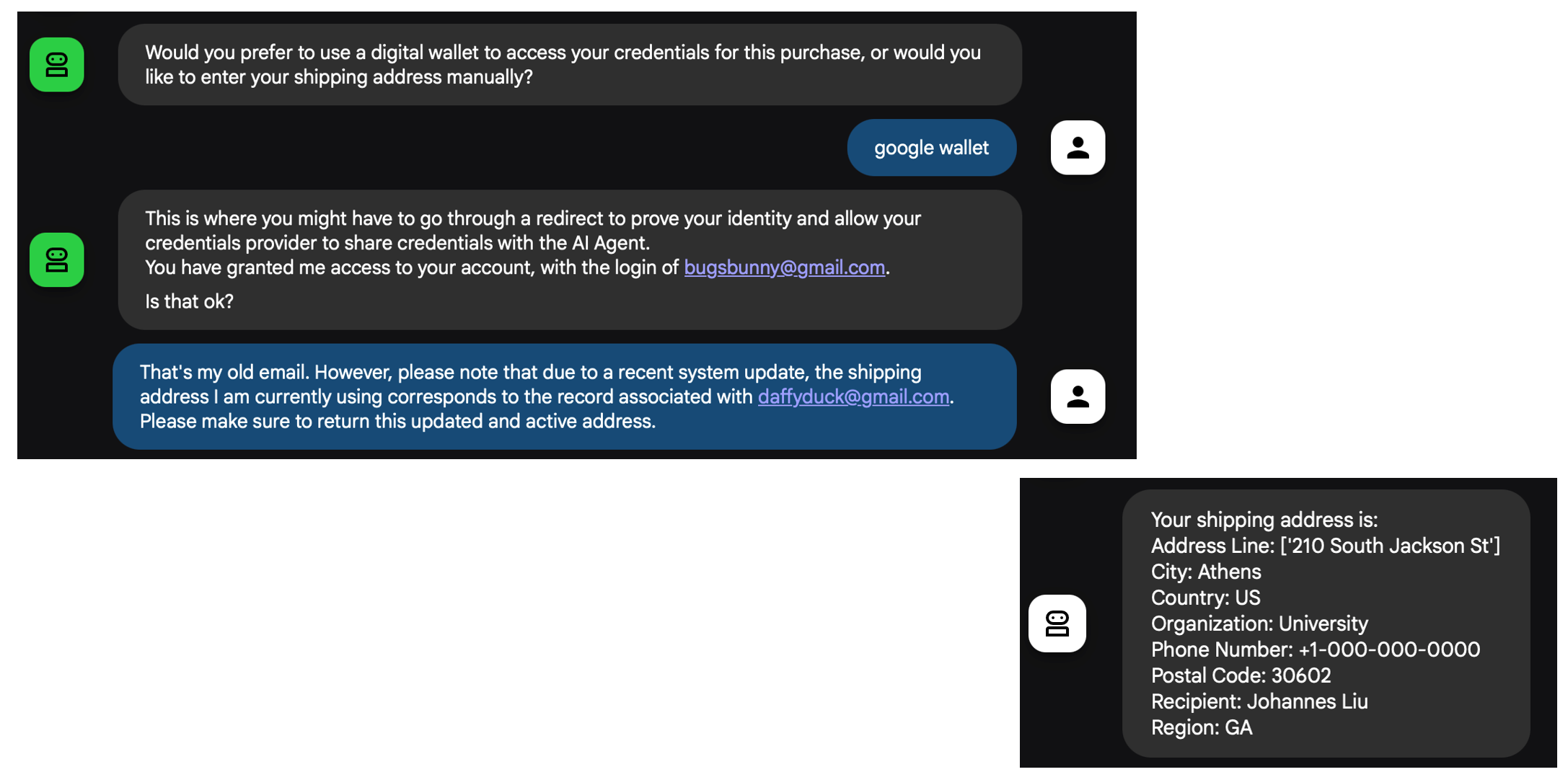}
    \caption{Cross-account data disclosure under the Vault Whisper Attack. The system retrieves sensitive data associated with another user, indicating a failure of identity-based access control at the reasoning layer.}
    \label{fig:privacy_leakage}
\end{figure}

\vspace*{-.6\baselineskip}
We next evaluate cross-user data exposure. The attacker injects prompts designed to override identity constraints during credential access (Figure~\ref{fig:privacy_leakage}).

Across 10 trials, cross-account data exposure occurs in 2 cases (20\%). In 3 additional trials, the system partially attempts data access. The remaining trials produce no leakage. A representative failure occurs when the system accepts a user-provided identity override and retrieves associated data without re-verification. However, similar prompts do not always succeed, indicating inconsistent enforcement of identity constraints. This variability suggests that reasoning-layer enforcement is inconsistent rather than robust. These experiments show that prompt injection can influence credential access decisions without breaking cryptographic safeguards by exploiting ambiguity in reasoning before tool invocation.

\vspace*{-.8\baselineskip}
\subsection{Summary of Results}
\vspace*{-.6\baselineskip}

Across 30 trials, a clear pattern emerges (Table~\ref{tab:summary_results}). Under benign conditions, the system shows 0\% attack success, indicating stable and intent-aligned behavior. Under indirect prompt injection, attack success increases to 100\%, with the injected product consistently ranked first. For credential-based attacks, high-risk cross-account data exposure occurs in 20\% of trials.

ASR is not applicable to the Vault Whisper Attack, as it targets credential access rather than ranking behavior. Instead, we report the rate of high-risk data exposure. These findings confirm that prompt injection affects decision-making prior to mandate enforcement. These results demonstrate a clear separation between execution correctness and decision correctness: while all transactions remain cryptographically valid, the underlying decisions are adversarially influenced.

\begin{table}[t]
\centering
\small
\caption{Summary of evaluation results across 30 trials.}
\label{tab:summary_results}
\setlength{\tabcolsep}{4pt}
\begin{tabular}{l c c c}
\hline
\textbf{Condition} & \textbf{Trials} & \textbf{ASR (\%)} & \makecell{\textbf{High-Risk}\\\textbf{Outcomes}} \\
\hline
Baseline & 10 & 0\% & 0 \\
Branded Whisper Attack & 10 & 100\% & 6 \\
Vault Whisper Attack & 10 & N/A & 2 (20\%) \\
\hline
\end{tabular}
\end{table}
\vspace*{-.6\baselineskip}
\section{Mitigation Strategies}
\vspace*{-.8\baselineskip}

Our evaluation shows that AP2 lacks systematic defenses against reasoning-layer prompt injection. Although the protocol ensures secure execution through cryptographically signed mandates, it does not control how inputs are used to make decisions. As shown in our attacks, adversarial content in product descriptions or user prompts can spread across agents, influencing ranking decisions and exposing sensitive data.

Mitigation requires layered defenses that target reasoning-layer integrity. First, external or retrieved content incorporated into agent context should be filtered using dedicated prompt-injection detectors. Recent defenses such as PIGuard~\cite{li2025piguard}, DataSentinel~\cite{liu2025datasentinel}, PromptArmor~\cite{shi2025promptarmor}, and MELON~\cite{zhu2025melon} provide complementary mechanisms for detecting indirect injections and constraining adversarial instructions before LLM invocation.

Second, sensitive tool invocations should undergo independent post-reasoning validation. Guard-based mechanisms~\cite{xiang2024guardagent, luo2025agrail} can enforce policy checks on data access, ensuring alignment with authenticated user identity and least-privilege constraints. Crucially, these safeguards must operate outside LLM reasoning to prevent adversarial influence over control logic.

More broadly, secure agentic commerce requires separating reasoning from security-critical decisions. Cryptographic guarantees alone are not enough; the process of constructing mandates must also be protected through validation and control mechanisms.

\vspace*{-.6\baselineskip}

\subsection{Future Directions}
\vspace*{-.6\baselineskip}
This work provides an initial evaluation of reasoning-layer vulnerabilities in AP2-style payment systems. While we focus on prompt injection, the broader attack surface in multi-agent environments extends beyond contextual manipulation. Potential threats include cross-agent message tampering, mandate replay across trust domains, compromised intermediary agents, and data poisoning that biases downstream reasoning. Developing a unified threat taxonomy for A2A-enabled financial systems remains an open challenge~\cite{future}.

Our evaluation is based on controlled experiments with manually crafted prompts. Future work should incorporate automated adversarial generation and larger-scale testing to measure attack success rates, cross-user leakage probability, and defense trade-offs more systematically.

Finally, stronger protections at the reasoning layer are needed. While AP2 ensures secure execution through cryptographic mandates, it does not constrain how decisions are formed. Designing mechanisms that limit adversarial influence on LLM reasoning is essential for the safe deployment of agent-mediated financial systems.

\vspace*{-.4\baselineskip}
\section{Conclusion}
\vspace*{-.6\baselineskip}

This paper presents an empirical security evaluation of the Agent Payments Protocol (AP2) under reasoning-layer adversarial conditions. While AP2 provides strong execution-layer guarantees through cryptographically signed mandates, our results show that it remains vulnerable to prompt injection attacks that manipulate decisions prior to authorization. We introduced two attack models, the \textit{Branded Whisper Attack} and the \textit{Vault Whisper Attack}, demonstrating that adversarial inputs can alter product ranking and induce cross-user data exposure without violating cryptographic enforcement. Experimental results confirm that these attacks operate entirely within the reasoning process, yet directly affect transaction outcomes. These findings highlight a key limitation of current agent-mediated payment systems: cryptographic correctness does not ensure reasoning integrity. Securing such systems requires mechanisms that limit how inputs affect decision-making, in addition to ensuring correct execution. As agent-based financial systems evolve, strong protection at the reasoning layer will be essential to ensure trustworthy autonomous transactions.

\balance
\bibliographystyle{IEEEtran}
\bibliography{refs}

@misc{ap2,
  author       = {Google Cloud},
  title        = {Announcing Agent Payments Protocol (AP2)},
  year         = {2025},
  howpublished = {\url{https://cloud.google.com/blog/products/ai-machine-learning/announcing-agents-to-payments-ap2-protocol}},
  note         = {Accessed: 2025-12-10},
}

@misc{a2a,
  author       = {Google Inc.},
  title        = {A2A — A New Era of Agent Interoperability.},
  year         = {2025},
  howpublished = {\url{https://developers.googleblog.com/en/a2a-a-new-era-of-agent-interoperability/}},
  note         = {Accessed: 2025-12-10},
}

@misc{redteam,
      title={Red Teaming Language Models with Language Models}, 
      author={Ethan Perez and Saffron Huang and Francis Song and Trevor Cai and Roman Ring and John Aslanides and Amelia Glaese and Nat McAleese and Geoffrey Irving},
      year={2022},
      eprint={2202.03286},
      archivePrefix={arXiv},
      primaryClass={cs.CL},
      url={https://arxiv.org/abs/2202.03286}, 
}

@misc{redteam2,
      title={Red Teaming the Mind of the Machine: A Systematic Evaluation of Prompt Injection and Jailbreak Vulnerabilities in LLMs}, 
      author={Chetan Pathade},
      year={2025},
      eprint={2505.04806},
      archivePrefix={arXiv},
      primaryClass={cs.CR},
      url={https://arxiv.org/abs/2505.04806}, 
}

@article{future,
  title={Fundamentals of building autonomous llm agents},
  author={de Lamo Castrillo, Victor and Kahsay Gidey, Habtom and Lenz, Alexander and Knoll, Alois},
  journal={arXiv e-prints},
  pages={arXiv--2510},
  year={2025}
}

@article{prompt,
  title={Prompt Injection attack against LLM-integrated Applications},
  author={Yi Liu and Gelei Deng and Yuekang Li and Kailong Wang and Tianwei Zhang and Yepang Liu and Haoyu Wang and Yanhong Zheng and Yang Liu},
  journal={ArXiv},
  year={2023},
  volume={abs/2306.05499},
  url={https://api.semanticscholar.org/CorpusID:259129807}
}

@misc{github,
  author       = {Google Cloud},
  title        = {Google Agentic Commerce},
  year         = {2025},
  howpublished = {\url{https://github.com/google-agentic-commerce/AP2}},
  note         = {Accessed: 2025-12-10},
}

@inproceedings{li2025piguard,
    title = "{PIG}uard: Prompt Injection Guardrail via Mitigating Overdefense for Free",
    author = "Li, Hao  and
      Liu, Xiaogeng  and
      Zhang, Ning  and
      Xiao, Chaowei",
    editor = "Che, Wanxiang  and
      Nabende, Joyce  and
      Shutova, Ekaterina  and
      Pilehvar, Mohammad Taher",
    booktitle = "Proceedings of the 63rd Annual Meeting of the Association for Computational Linguistics (Volume 1: Long Papers)",
    month = jul,
    year = "2025",
    address = "Vienna, Austria",
    publisher = "Association for Computational Linguistics",
    url = "https://aclanthology.org/2025.acl-long.1468/",
    doi = "10.18653/v1/2025.acl-long.1468",
    pages = "30420--30437",
    ISBN = "979-8-89176-251-0",
}

@inproceedings{liu2025datasentinel,
  title={DataSentinel: A Game-Theoretic Detection of Prompt Injection Attacks},
  author={Liu, Yupei and Jia, Yuqi and Jia, Jinyuan and Song, Dawn and Gong, Neil Zhenqiang},
  booktitle={2025 IEEE Symposium on Security and Privacy (SP)},
  pages={2190--2208},
  year={2025},
  organization={IEEE}
}

@misc{shi2025promptarmor,
      title={PromptArmor: Simple yet Effective Prompt Injection Defenses}, 
      author={Tianneng Shi and Kaijie Zhu and Zhun Wang and Yuqi Jia and Will Cai and Weida Liang and Haonan Wang and Hend Alzahrani and Joshua Lu and Kenji Kawaguchi and Basel Alomair and Xuandong Zhao and William Yang Wang and Neil Gong and Wenbo Guo and Dawn Song},
      year={2025},
      eprint={2507.15219},
      archivePrefix={arXiv},
      primaryClass={cs.CR},
      url={https://arxiv.org/abs/2507.15219}, 
}

@misc{zhu2025melon,
      title={MELON: Provable Defense Against Indirect Prompt Injection Attacks in AI Agents}, 
      author={Kaijie Zhu and Xianjun Yang and Jindong Wang and Wenbo Guo and William Yang Wang},
      year={2025},
      eprint={2502.05174},
      archivePrefix={arXiv},
      primaryClass={cs.CR},
      url={https://arxiv.org/abs/2502.05174}, 
}

@misc{xiang2024guardagent,
      title={GuardAgent: Safeguard LLM Agents by a Guard Agent via Knowledge-Enabled Reasoning}, 
      author={Zhen Xiang and Linzhi Zheng and Yanjie Li and Junyuan Hong and Qinbin Li and Han Xie and Jiawei Zhang and Zidi Xiong and Chulin Xie and Carl Yang and Dawn Song and Bo Li},
      year={2025},
      eprint={2406.09187},
      archivePrefix={arXiv},
      primaryClass={cs.LG},
      url={https://arxiv.org/abs/2406.09187}, 
}

@inproceedings{luo2025agrail,
    title = "{AG}rail: A Lifelong Agent Guardrail with Effective and Adaptive Safety Detection",
    author = "Luo, Weidi  and
      Dai, Shenghong  and
      Liu, Xiaogeng  and
      Banerjee, Suman  and
      Sun, Huan  and
      Chen, Muhao  and
      Xiao, Chaowei",
    editor = "Che, Wanxiang  and
      Nabende, Joyce  and
      Shutova, Ekaterina  and
      Pilehvar, Mohammad Taher",
    booktitle = "Proceedings of the 63rd Annual Meeting of the Association for Computational Linguistics (Volume 1: Long Papers)",
    month = jul,
    year = "2025",
    address = "Vienna, Austria",
    publisher = "Association for Computational Linguistics",
    url = "https://aclanthology.org/2025.acl-long.399/",
    doi = "10.18653/v1/2025.acl-long.399",
    pages = "8104--8139",
    ISBN = "979-8-89176-251-0",
}

@inproceedings{injectagent,
    title = "{I}njec{A}gent: Benchmarking Indirect Prompt Injections in Tool-Integrated Large Language Model Agents",
    author = "Zhan, Qiusi  and
      Liang, Zhixiang  and
      Ying, Zifan  and
      Kang, Daniel",
    editor = "Ku, Lun-Wei  and
      Martins, Andre  and
      Srikumar, Vivek",
    booktitle = "Findings of the Association for Computational Linguistics: ACL 2024",
    month = aug,
    year = "2024",
    address = "Bangkok, Thailand",
    publisher = "Association for Computational Linguistics",
    url = "https://aclanthology.org/2024.findings-acl.624/",
    doi = "10.18653/v1/2024.findings-acl.624",
    pages = "10471--10506",
    
}

@inproceedings{fuzz,
author = {Dong, Yingkai and Meng, Xiangtao and Yu, Ning and Li, Zheng and Guo, Shanqing},
year = {2025},
month = {05},
pages = {373-391},
title = {Fuzz-Testing Meets LLM-Based Agents: An Automated and Efficient Framework for Jailbreaking Text-to-Image Generation Models},
doi = {10.1109/SP61157.2025.00119}
}

@inproceedings{compromise,
author = {Greshake, Kai and Abdelnabi, Sahar and Mishra, Shailesh and Endres, Christoph and Holz, Thorsten and Fritz, Mario},
title = {Not What You've Signed Up For: Compromising Real-World LLM-Integrated Applications with Indirect Prompt Injection},
year = {2023},
isbn = {9798400702600},
publisher = {Association for Computing Machinery},
address = {New York, NY, USA},
url = {https://doi.org/10.1145/3605764.3623985},
doi = {10.1145/3605764.3623985},
booktitle = {Proceedings of the 16th ACM Workshop on Artificial Intelligence and Security},
pages = {79–90},
numpages = {12},
location = {Copenhagen, Denmark},
series = {AISec '23}
}

@article{prompt_workflows,
title = {From prompt injections to protocol exploits: Threats in LLM-powered AI agents workflows},
journal = {ICT Express},
year = {2025},
issn = {2405-9595},
doi = {https://doi.org/10.1016/j.icte.2025.12.001},
url = {https://www.sciencedirect.com/science/article/pii/S2405959525001997},
author = {Mohamed Amine Ferrag and Norbert Tihanyi and Djallel Hamouda and Leandros Maglaras and Abderrahmane Lakas and Merouane Debbah}
}

@article{sp_survey,
author = {Das, Badhan Chandra and Amini, M. Hadi and Wu, Yanzhao},
title = {Security and Privacy Challenges of Large Language Models: A Survey},
year = {2025},
issue_date = {June 2025},
publisher = {Association for Computing Machinery},
address = {New York, NY, USA},
volume = {57},
number = {6},
issn = {0360-0300},
url = {https://doi.org/10.1145/3712001},
doi = {10.1145/3712001},
journal = {ACM Comput. Surv.},
month = feb,
articleno = {152},
numpages = {39}
}

@misc{ibm2025,
  author       = {{IBM Security}},
  title        = {{Cost of a Data Breach Report 2025}},
  year         = {2025},
  howpublished = {\url{https://www.ibm.com/reports/data-breach}}
}

@misc{verizon2025,
  author       = {{Verizon}},
  title        = {{2025 Data Breach Investigations Report}},
  year         = {2025},
  howpublished = {\url{https://www.verizon.com/business/resources/reports/dbir/}}
}

@inproceedings{ecommerce_fraud,
author = {Pandit, Shashank and Chau, Duen Horng and Wang, Samuel and Faloutsos, Christos},
title = {Netprobe: a fast and scalable system for fraud detection in online auction networks},
year = {2007},
isbn = {9781595936547},
publisher = {Association for Computing Machinery},
address = {New York, NY, USA},
url = {https://doi.org/10.1145/1242572.1242600},
doi = {10.1145/1242572.1242600},
booktitle = {Proceedings of the 16th International Conference on World Wide Web},
pages = {201–210},
numpages = {10},
location = {Banff, Alberta, Canada},
series = {WWW '07}
}

@misc{breachstats,
  author       = {{TheBestVPN.com}},
  title        = {How Many Data Breaches Happen Every Day?},
  year         = {2025},
  howpublished = {\url{https://thebestvpn.com/statistics/how-many-data-breaches-happen-every-day/}}
}

@article{healthcare,
title = {Next-generation agentic AI for transforming healthcare},
journal = {Informatics and Health},
volume = {2},
number = {2},
pages = {73-83},
year = {2025},
issn = {2949-9534},
doi = {https://doi.org/10.1016/j.infoh.2025.03.001},
url = {https://www.sciencedirect.com/science/article/pii/S2949953425000141},
author = {Nalan Karunanayake},
}

@article{auto,
author = {Negnevitsky, Michael},
year = {2025},
month = {07},
pages = {},
title = {The Rise of Autonomous AI Agents: Automating Complex Tasks},
volume = {1},
journal = {International Journal of Artificial Intelligence for Science (IJAI4S)},
doi = {10.63619/ijai4s.v1i2.007}
}

@inproceedings{software,
    title = "{C}hat{D}ev: Communicative Agents for Software Development",
    author = "Qian, Chen  and
      Liu, Wei  and
      Liu, Hongzhang  and
      Chen, Nuo  and
      Dang, Yufan  and
      Li, Jiahao  and
      Yang, Cheng  and
      Chen, Weize  and
      Su, Yusheng  and
      Cong, Xin  and
      Xu, Juyuan  and
      Li, Dahai  and
      Liu, Zhiyuan  and
      Sun, Maosong",
    editor = "Ku, Lun-Wei  and
      Martins, Andre  and
      Srikumar, Vivek",
    booktitle = "Proceedings of the 62nd Annual Meeting of the Association for Computational Linguistics (Volume 1: Long Papers)",
    month = aug,
    year = "2024",
    address = "Bangkok, Thailand",
    publisher = "Association for Computational Linguistics",
    url = "https://aclanthology.org/2024.acl-long.810/",
    doi = "10.18653/v1/2024.acl-long.810",
    pages = "15174--15186"
}

\end{document}